\documentclass[12pt]{article}
\usepackage{graphics}
\usepackage{epsfig} 
\usepackage{hyperref}
\usepackage{graphics}
\usepackage{color}      
\usepackage{graphicx}

\title{\bf Stochastic Resonance of a Flexible Chain Crossing over a Barrier  }
\author{ Mesfin Asfaw$^{a}$ and Wokyung Sung$^{b}$ \\
     $^{a}$Asian Pacific Center for Theoretical Physics,  Pohang 790-784,\\ South Korea \\
$^{b}$Department of Physics and PCTP, Pohang \\University of Science and Technology, Pohang 790-784, South Korea}
 
\date{Received: date / Revised version: date}

\begin{document}
\maketitle

\begin{abstract}
We study the stochastic resonance (SR) of a flexible polymer surmounting
a bistable-potential barrier. Due to the flexibility that can
enhance crossing rate and change chain conformations at 
the barrier, the SR behaviors manifest many features of an entropic
SR of a new kind, such as the power amplification peaks at optimal
chain lengths and elastic constants as well as the optimal noise
strengths. The pronounced peaks that emerge depending on the chain
lengths and conformation states suggest novel means of manipulating
biopolymers, such as efficient separation methods, within undulating channels.
\end{abstract}
 
\maketitle
The stochastic resonance (SR) is a counter-intuitive phenomenon, where
background noise can be instrumental in enhancing coherence and resonance
of a nonlinear system to a small periodic signal \cite{b4,b5}. It occurs when noise-induced hopping events synchronize with the signal, which itself is not sufficient to drive the system to cross over the barrier.  Since the pioneering work of Benz et al.
on recurrence of ice ages \cite{b3}, the idea of SR has been extended to a
vast range of phenomena encompassing signal and information processing, medicine, and biological systems [4-6]. Recently a number of workers reported the new mechanism of entropic stochastic resonance for the Brownian particle which is confined in a narrow space and thus subject to free energy rather than purely energetic potential \cite{b9}. For
nano-scaled soft matter /biological systems the ubiquitous entropic effects due to confinement
can play an important role in noise-induced resonant effects as recently have been shown in the studies of SR in ion channels \cite{b10,b11}.

As interconnected, flexible systems they are, biopolymers manifest interesting
cooperative dynamics in certain confined environments, external fields
and noises. Its cooperative dynamics is important, not only in understanding
how such a system self-organizes its flexible degrees of freedom, but
also in a multitude of single molecule biophysics applications such DNA separation,
and biopolymer sequencing, etc. Nature can  utilize the ambient noises
(fluctuations) of various kinds in such biological soft-condensed matter to facilitate the barrier
crossing seemingly difficult to surmount, typically assisted by conformational
changes. 

The study of dynamics of a polymer surmounting a potential barrier along with the associated SR provides a basic paradigm in which to understand the self-organization and cooperativity induced by the chain flexibility and fluctuations. The potential force on the chain can be traded with an entropic force caused by confining geometry as noted by \cite{a1}. This  make it possible to study   the dynamics under an external potential by probing the equivalent dynamics within a channel with the cross section modulating over mesoscopic or macroscopic length scale that can be fabricated \cite{a2}. For the characteristic dimension of the double well potential much longer than the chain contour length, the crossing  (Kramers) rate  for a flexible chain was found to be always higher than expected for the rigid globule it becomes in the limit of infinitely high elastic constant \cite{b12,e2}. Also  the rate is found to  depend on the conformation, coiled or stretched, which the flexible chain takes at the energy barrier. These behaviors  suggest the emergence of an entropic SR of a new kind originating from chain's intrinsic flexibility, which we study in this Letter.

Based on the crossing rates of such chain given in \cite{e2}, whose validity is confirmed by simulations \cite{a3}, we recast the many-body  dynamics of segments to that of the  center of mass (cm) of the chain under an effective potential or free energy. To identify the most salient features of this SR, we analyze the power amplification factor via a linear response theory.
The power amplification factors manifest peaks as the typical signature of SR at optimal noise
strengths, but also, remarkably, at optimal chain lengths that are different
depending on the chain flexibility and the conformations at the barrier or a narrow constriction in a channel. Therefore the SR suggests novel possibilities of understanding the biopolymers living processes, and manipulating single biopolymers utilizing its flexibility and sorting them depending on their lengths \cite{b15}. Earlier studies on SR of linearly coupled chains have mostly focused on the other extreme, i.e., the chains much longer than the width of the barrier. Here the barrier crossing is facilitated by a mode of excitation called "kink-anti kink pair", which yields the crossing rate \cite{b16} and SR \cite{b17,b18} much different from discussed above. Because the kink is a localized object in the long chain limit, the activation
energy is independent of the chain lengths and conformations, unlike in many
practical cases, with which we are concerned.

We consider that a linear, harmonic chain of N beads (monomers) undergoes a Brownian motion in three dimension. Assuming each bead has a friction coefficient $\gamma$, the dynamics of the $N$ beads ( $n$=1,2,3 . . . $N$) under a one-dimensional external potential $U(x)$ and thermal activation is governed by the Langevin equation,
\begin{equation}
\gamma{dx_{n}\over dt}=-k(2x_{n}-x_{n-1}-x_{n+1})-{\partial U(x_{n})\over \partial x_{n}}+  \xi_{n}(t)
\end{equation}
where the $k$ is the spring (elastic constant) of the chain, $\xi_{n}(t)$ is assumed to be Gaussian and white noise satisfying
\begin{equation}
\left\langle  \xi_{n}(t) \right\rangle =0,~~~\left\langle \xi_{n}(t)  \xi_{n}(t+\tau) \right\rangle=2D\gamma
\delta(\tau)
\end{equation}
with $D= k_{B}T$ is the strength of the thermal noise. The external potential
energy each bead experiences is given by
\begin{equation}
U(x)=-({\omega_{B}^{2} \over 2})x^{2}+{1\over 4}({\omega_{B}^{2} \over x_{m}^{2}})x^{4}
\end{equation} 
as shown in Fig. 1. The  positions $x= \pm x_{m}$ represent    
the two minima  of   the potential separated by a barrier of height $U_{B}= x_{m}^{2}\omega_{B}^{2}/4$ which  is  centered at $x=0$.
The parameter 
$\omega_{B}^{2}={1\over 2}\omega_{0}^{2}$ and  $\omega_{0}^{2}$ denote the potential curvatures of the barrier top and the well minima respectively.
\begin{figure}[ht]
\epsfig{file=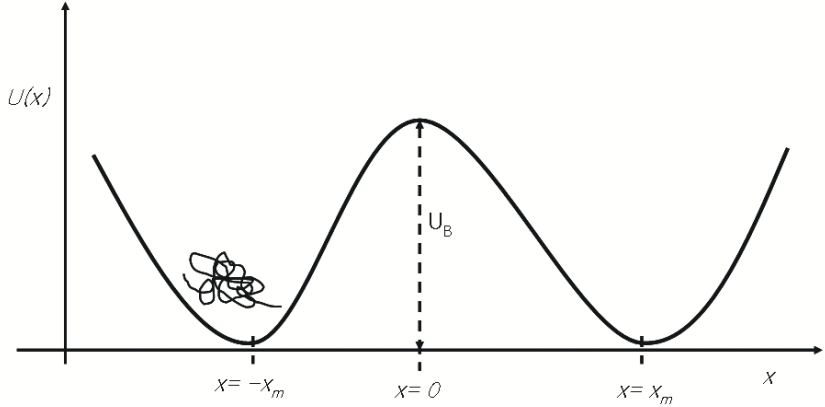,width=12cm}
\caption{Schematic diagram for a flexible polymer chain in a symmetric double-well potential $U(x)=-({\omega_{B}^{2} \over 2})x^{2}+{1\over 4}({\omega_{B}^{2} \over x_{m}^{2}})x^{4}$. The potential wells and the barrier top are located at $x=\pm x_{m}$ and  at $x=0$. We consider the case of the potential where $NU_B \gg D$, and $L \gg X_m $  so that the potential is designed to vary on a mesoscopic or macroscopic scale. }  
\end{figure}

We ask a fundamental question: How does the linearly coupled flexible system like a polymer chain in the double well cooperatively respond to a small time-periodic force in the presence of the ambient thermal noise? The determinant parameters are the noise strength (temperature), the chain length, and the elastic constant, given the double well potential parameters as fixed.  By such coupling the SR is enhanced above that of a single bead as already has been studied in earlier studies [17,18]. In the globular limit where the coupling (spring) constant is infinity, the polymer moves as a single Brownian particle (a rigid globule) in the bistable potential, whose SR behavior is well-known. Now turning on the internal degrees of freedom that give rise to the chain conformational flexibility and variability, how will the SR be affected? 

 We base our theoretical development of SR on previous theoretical results [13] for the rate of the chain crossing over the potential barrier by thermal activation in the absence of driving force,
which is found to be consistent with numerical simulations [13, 14]. The rate is given by
\begin{equation}
R={R_{0}} e^{-{\Delta F^{'}\over D}}
\end{equation}
 where
 \begin{equation}
R_{0}={\omega_{B}\omega_{0}\over 2\pi \gamma} e^{-{NU_{B}\over D}}.
\end{equation} Here the $R_0$ is the rate in the globular limit, $k\rightarrow \infty$. As the chain attains
the flexibility by decreasing $k$, the rate is enhanced by the factor $e^{-{\Delta F^{'}\over D}}$
where $\Delta F^{'}$ is the free energy change of the chain during the barrier crossing, which depends on temperature.

 It is because   the chain, a flexible extended object as it is, experiences an entropy increase in surpassing the potential, i.e., from the confining well to the unstable barrier top, 
 thereby reducing the activation energy barrier \cite{e2}. $\Delta F^{'}$ depends on $D$ as shown below. Depending upon the chain length $N$ and the rescaled spring constant ${\bar k}=k/\omega_{B}^{2}$, the flexible chain takes either  the coiled or stretched conformation at the barrier top (at the transition state).  For the regime of parameters satisfying $q>0$ where 
 \begin{equation}
 q={k \pi\over N\omega_{B}^{2}}-1
 \end{equation}
  the chain retains the coiled conformation, the free energy change is expressed by 
 \begin{eqnarray}
\Delta F^{'}= -D\ln \left[\left({\omega_{B}^{2}\over \omega_{0}^{2}}\right]^{1/4} \left({sinh(N\sqrt{\omega_{0}^{2}/k})\over sin( N\sqrt{\omega_{B}^{2}/k})}\right)^{1/2}f(\alpha)\right)=-T\Delta S_{c}
\end{eqnarray}
and $ 
f(\alpha)=\sqrt{\alpha \over 2\pi}\int_{-\infty}^{\infty} dQ e^{-(\alpha/2)Q^{2}-{3/8}Q^{4}}$
with $ 
\alpha=2q\left({U_{B}\over D}\right)^{1\over 2}$. It should be noted that $\Delta F^{'}$  is given solely  from the entropy change $\Delta S_{c}$ mentioned above. On the other hand, when $q<0$, the chain takes a stretched conformation at the barrier top, thereby reducing the energy barrier by the amount 
 \begin{eqnarray}
\Delta F^{'}=-N\omega_{B}^{2}x_{m}^{2}q^{2}/6- D \ln\left(\sqrt{2}\left({\tilde{\omega_{B}}\over \omega_{B}}\right)  \left({\tilde{\omega_{B}}^{2}\over \omega_{0}^{2}}\right)^{1/4} \left({sinh(N\sqrt{\omega_{0}^{2}/k})\over sin( N\sqrt{\tilde{\omega_{B}}^{2}/k})}\right)^{1/2}g(\alpha)\right)
\end{eqnarray}
for small values of $|q|$. In Eq. (9) $
g(\alpha)=\sqrt{\alpha \over 4\pi}e^{-\alpha^{2}/6}\int_{-\infty}^{\infty} dQ e^{-(\alpha/2)Q^{2}-{3/8}Q^{4}}
$
where $(\tilde{\omega_{B}})^2=(1-2|q|)\omega_{B}^2$. Eq (8) includes the decrease in the internal energy due to stretching ( the first term) as well as the entropy increase    (the second term) in crossing the barrier in stretched state. 
The transition from coiled to stretched conformation at the barrier are incurred as the chain length increases above $N_{c}$ or  the spring constant decreases below a critical value $k_{c}$ with the potential parameter fixed.

Driven by a force $F(t)= F_{0}\cos{(\Omega t)}$ on each bead, where $F_{0}$ and $\Omega$ are the amplitude and angular frequency,  the center of mass position $X$ of the chain, the reaction coordinate for the barrier crossing, evolves following the Langevin equation under an effective potential or free energy function $F(X)$
  \begin{equation}
N\gamma{{dX}\over dt}= -{\partial F(X)\over \partial X}+ NF_{0}cos{(\Omega t)}+\xi(t)
\end{equation} 
where  $\xi(t)$ represents   the random force characterized by $\left\langle  \xi(t) \right\rangle =0,~~~\left\langle \xi(t)  \xi(t+\tau) \right\rangle=2DN\gamma \delta(\tau)$.

It is reasonable to assume  the free energy   $F(X)$ associated with the cm position  $X$ takes the form  similar to Eq. (3), 
\begin{equation}
F(X)=-{\Omega_{B}^{2} \over 2}X^{2}+{\Omega_{B}^{2} \over 4x_{m}^{2}}X^{4}
\end{equation}
with the  barrier curvature  $\Omega_{B}^{2}$ and  height $\Delta F={1\over 4}\Omega_{B}^{2}x_{m}^{2}$.
The Kramer rate for the cm regarded effectively as a single Brownian particle is 
\begin{equation}
R={\Omega_{B}\Omega_{0}\over 2\pi N \gamma} e^{-{\Delta F\over D }}
\end{equation}
which is equated with Eq. (4)   to yield 
\begin{equation}
\Delta  F\cong NU_{B}+\Delta F^{'}
\end{equation}
and thus
\begin{equation}
\Omega_{B}=\sqrt{N}\omega_{B}\left(1+{4\Delta F^{'}\over N\omega_{B}^{2}x_{m}^2}\right)^{1\over 2},
\end{equation}
both of which depend on $N$ as well as the $D$ via $\Delta F^{'}$. The free energy function, rather than purely energetic potential, under which the particle representative of the chain is moving, signals an occurrence of an entropic stochastic resonance \cite{b9}. 

How can we realize the dynamics of the chain under the one-dimensional  free energy $F(X)$ in a manageable experiment? To this end we convert the dynamics to that of the chain flowing within the channel of undulating cross section $A(X)$. Since we envision the chain as a single particle moving subject to $F(X)$, such a conversion is possible if  $F(X)= -D\ln A(X)$, or 
$A(X)=Ao\exp(-F(X)/D$, where $A_0$ is the cross section at the point where $F(X)=0$. Therefore the crossing dynamics and the associated SR can be achieved for a chain within such a fluidic  channel subject to an external periodic forcing, for example a single stranded DNA within a channel subject to AC field. 

To gain understanding of the major salient features of the distinctive nature
of the polymer SR, it suffices to consider the linear response of the chain cm to the small
driving forces. In response to the weak driving force, the average cm position
is given by
 \begin{equation}
\overline{X(t)}=NF_{0}|{\bar \chi}(\Omega)|\cos{(\Omega t-\phi)}
\end{equation} 
     where ${\bar \chi}(\Omega)$ is the Fourier transform of the response function \cite{b20} given by 
    $
     {\bar \chi}(\Omega)={\bar \chi}^{'}(\Omega)+i{\bar \chi}^{''}(\Omega)={\left\langle X^2\right\rangle \over D}{2R\over 2R+i \Omega}
     $
     and $\phi$ is the phase delay given by $\tan^{-1}{{\bar \chi^{''}}(\Omega) \over {\bar \chi^{'}}(\Omega)}$ while the coherence of the system is measured by the real part of ${\bar \chi}(\Omega)$, the SR intensity is quantified by the power amplification factor (the ratio of power stored in the response of the system to the power of the
driving force with frequency):
 \begin{equation}
\eta=|{\bar \chi}(\Omega)|^{2}=\left({ \left\langle X^2\right\rangle\over D}\right)^2{4R^2(D)\over 4R^2(D)+\Omega^2}
\end{equation}
where 
$
\left\langle X^2\right\rangle={\int X^{2}e^{-{F(X)\over D}}dX / \int e^{-{F(X)\over D}}dX}.
$

The $\eta$ indeed shows  nonmonotonic noise strength dependence manifesting a peak at an optimal noise strength, 
 which is the typical  signature of  SR.  Via dependence of the rate $R$ and $\left\langle X^2\right\rangle$ on the conformational variability, the SR for the flexible polymer manifests  distinctive dependence on the noise strength and additional dependence on the chain length, characteristic of a polymer entropic SR.   Below  we  quantitatively discuss the power amplification factor $\eta$  for the polymer conformation both below and above coil-to-stretch transition. We introduce  dimensionless parameters:  ${\bar D}=D/ \omega_{B}^{2}x_{m}^{2}$, ${\bar \Omega}=\gamma \Omega / \omega_{B}^{2}$ and ${\bar k}=k/\omega^{2}$. From now on all the quantities are rescaled (dimensionless) so   the bars will be dropped.  
\begin{figure}[ht]
\centering
{
    \includegraphics[width=7cm]{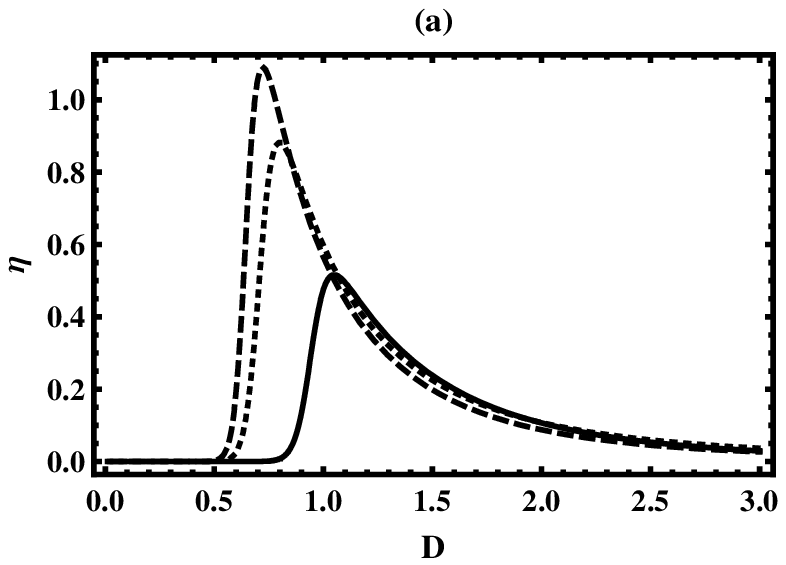}}
\hspace{1cm}
{
    \includegraphics[width=7cm]{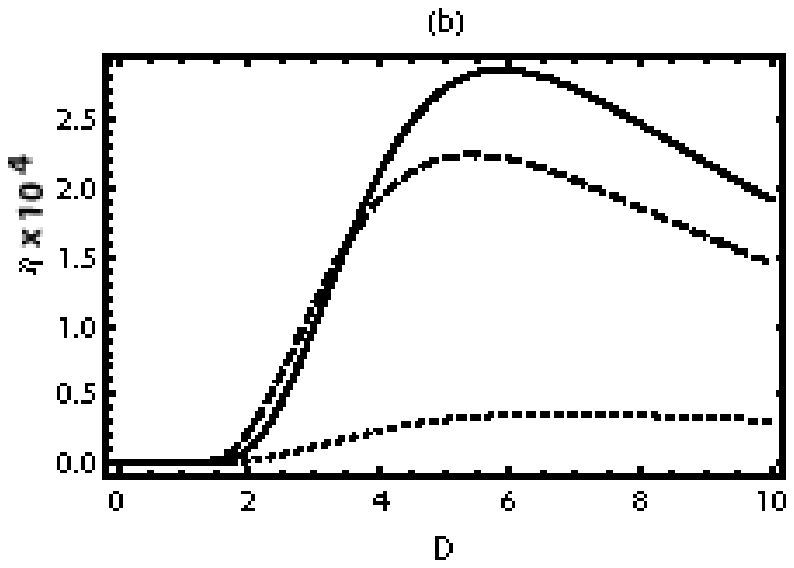}
}
\caption{(a) $\eta$ versus  $D$ for $\Omega=10^{-5}$. (b) $\eta \times 10^4$  versus  $D$ for $\Omega=1$. The solid line stands  for stretched  state ($N=50$), the  dashed line for coiled state ( $N=30$) at the barrier top and dotted line for the globule  ($N=30$) for parameter value  $k=200$ for which $N_{C}=44.4$ is the critical length for  the  coil-to-stretch transition.  } 
\label{fig:sub} 
\end{figure}

 Let us first calculate $\eta$ as a function of rescaled noise strength
$D$ for the rescaled spring constant  $k=200$
 which can be   reasonable for a real flexible polymer  if $\omega_{B}^{2}$ is chosen properly.   The Figures 2a and 2b  depict the rescaled   $\eta$ as a function of noise strength $D$ for $\Omega=10^{-5}$ and $\Omega=1$, respectively for various transition  states of polymer conformation; the coiled state with $N= 30$ and the stretched state with $N=50$, both of which are compared to globular state (the case of  infinite $k$) with $N= 30$. Hereafter in the figures, the coiled, stretched and globular states are depicted as dashed, solid  and dotted lines.
 As shown in the Figs. 2a and 2b, for all cases, the $\eta$ reach their  maximum value $\eta_{R}$  at certain optimal values of $D$, denoted by $D_{R}$. When  the chain is in coiled  state, $\eta_{R}$ is much larger  that of globular state. This signifies that its intrinsic flexibility  facilitating faster crossing allows the polymer chain to respond in a more cooperative and coherent manner to the external signal.
 
  For very small $\Omega$, as depicted in Fig. 2a  the optimal noise strength corresponding to the peaks satisfy  $D_{RC}<D_{RG}<D_{RS}$, where the subscripts $C$, $G$ and $S$ denote the three conformational states. To understand this, we note that the time scale matching condition between the mean crossing time and half of the period of oscillation, ${1\over R(D_{R})}={\pi\over \Omega}$, which provides a reasonable approximation for the resonant condition. With 
  $R(D_{R})\approx 3\times 10^{-6}$ given thus, and from the  $R$ obtained  from Eqs. (7) and (8) (see Fig. 2a), the above inequality indeed holds. The  figure 2a  shows that $\eta_{RC}>\eta_{RG}>\eta_{RS}$ which can be understood as follows.
  At very low frequents, the $\eta$, Eq. (15) is given by $
  \eta\approx|{\bar \chi}(0)|^{2}=\left({\left\langle X^2\right\rangle\over D}\right)^{2}
  $ 
  which implies that the resonance mechanism is governed by chain static susceptibility 
   $|{\bar \chi}(0)|= {\left\langle X^2\right\rangle\over D}$ of the cm in response to a constant  force. Finding the static susceptibility for the $D_{R}$ of each transition state $(D_{RC}<D_{RG}<D_{RS})$, one indeed affirms the inequality for the $\eta_{R}$.
  
  \begin{figure}[ht]
\centering
{
    \includegraphics[width=7cm]{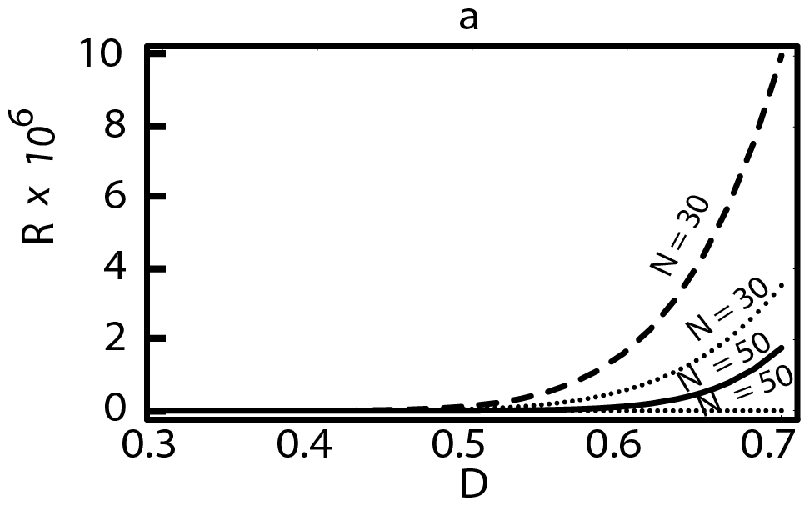}}
\hspace{1cm}
{
    \includegraphics[width=7cm]{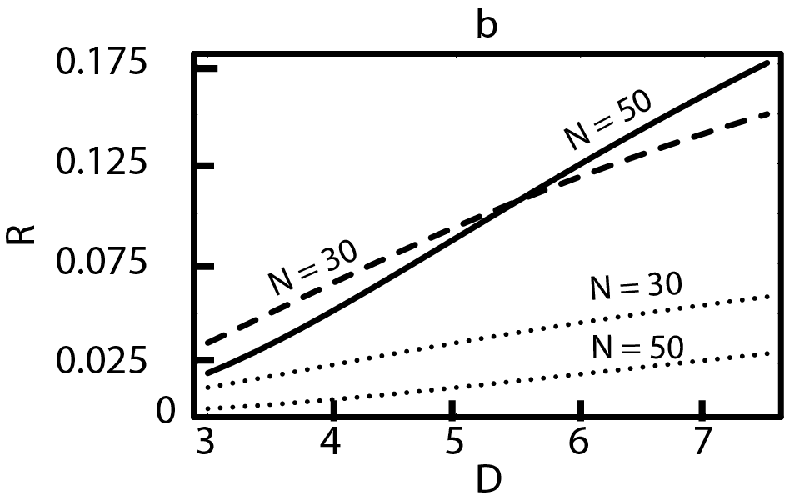}
}
\caption{ (a) Polymer crossing rate $R$ versus (times $10^6$)  $D$ for small values of $D$.  The dotted, dashed and solid lines stand for globular ($N=30$ and $N=50$), coiled ($N=30$) and stretched  state ($N=50$) at the top of the barrier.
 (b)  Polymer crossing rate $R$ versus   $D$ for large values of $D$.  } 
\label{fig:sub} 
\end{figure}
  
   The situation is different for large driving frequency where the resonance
conditions are met for larger $D_R$ (see Fig. 2b). While the inequality for $D_R$ is unchanged, that of $\eta_R$ is not; $\eta_R$ for stretched state is highest. This is understandable if we note for large $\Omega$ the $\eta$ is largely governed by the rate, rather than the static susceptibility. As the chain length
becomes longer than the critical value for this large $D$, this flexible chain
can extend to lower the free energy barrier and enhance the crossing rate, as
shown in Fig. 2b. This results in the stretched conformation enhancing SR (solid line) than the coiled or globular states at these noise strengths as the Eq.(20) indicates. This
means that the chain capable of the conformational transition can utilize such
flexibility to enhance the cooperatively and coherence to external influences. 
\begin{figure}[ht]
\centering
{
    \includegraphics[width=7cm]{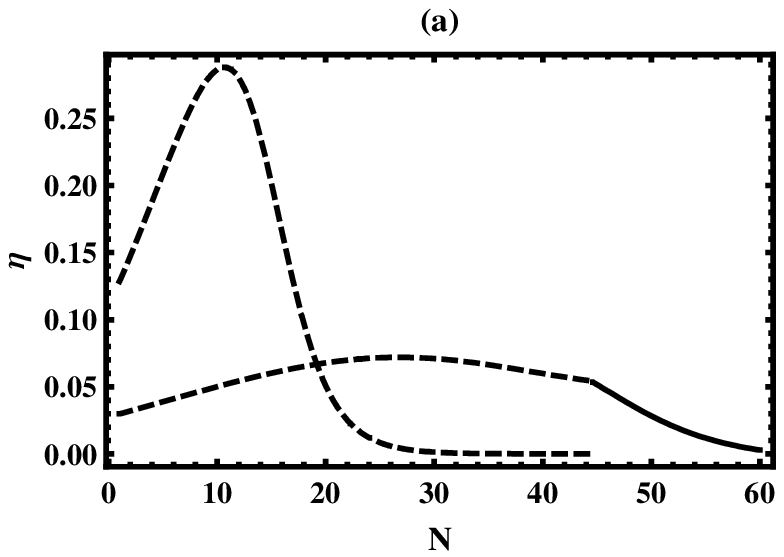}}
\hspace{1cm}
{
    \includegraphics[width=7cm]{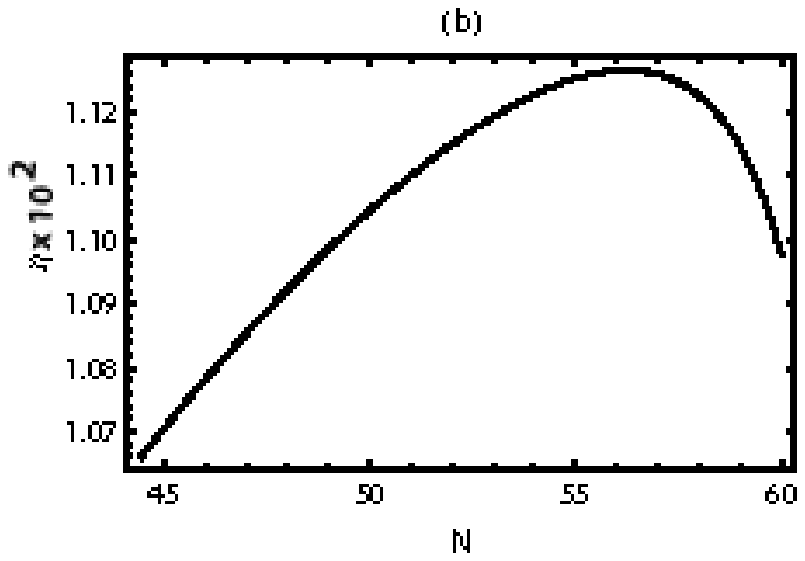}
}
\caption{(a)  $\eta$   as a function of $N$ for noise intensity $D=1$ (the top curve) and $D=2$ (lower curve). (b)  $\eta$ (times $10^{2}$)   as a function of $N$ for noise intensity $D=4$. 
 Other parameters are fixed as  $\Omega=0.01$ and $k/ \omega_{B}^{2}=200$.  } 
\label{fig:sub} 
\end{figure}

Another aspect of this entropic SR is the phenomenon above is shown in the Figures 4a and 4b which  depict   $\eta$  a function of chain length  with $k=200$ and  $\Omega=10^{-2}$ for three different noise strengths, $D=1$, $2$, and $4$.  For $D=1$, the chain manifest the response of coiled conformation for all values of  $N$ which are mostly smaller than $N_{c}=44.4$. For $D=2$, the chain responses in coiled conformation for $N<N_c$ but responses in stretched conformation for $N>N_c$.  For $D=4$ the chain manifest smaller $\eta$ for all chain of length $N>N_c$ which are all  in stretched conformation at the barrier top. Each of these lines have a peak at a certain optimal  $N$ which is increasing with $D$. Even with the temperature fixed, as is usual in biological systems, a long chain by stretching can escape the barrier or the narrowest constriction in a most coherent and resonant way to an external signal.    
\begin{figure}[ht] 
\centering
{ 
    \includegraphics[width=7cm]{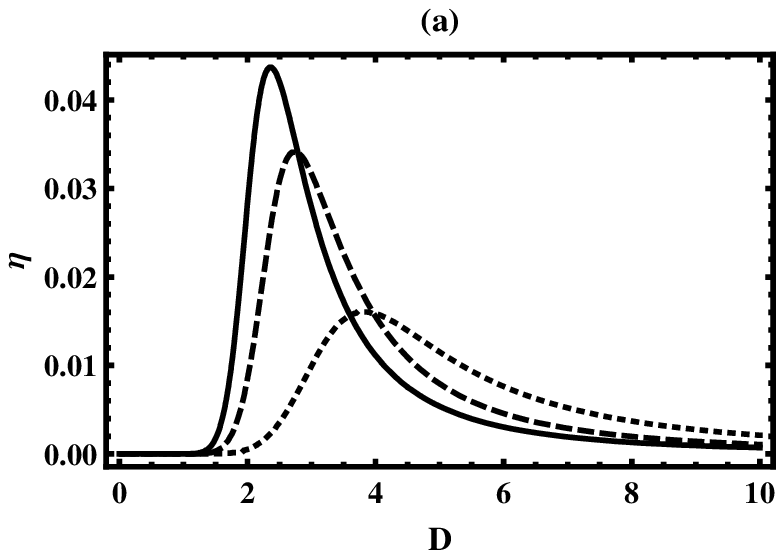}
}
\hspace{1cm}
{
    \includegraphics[width=7cm]{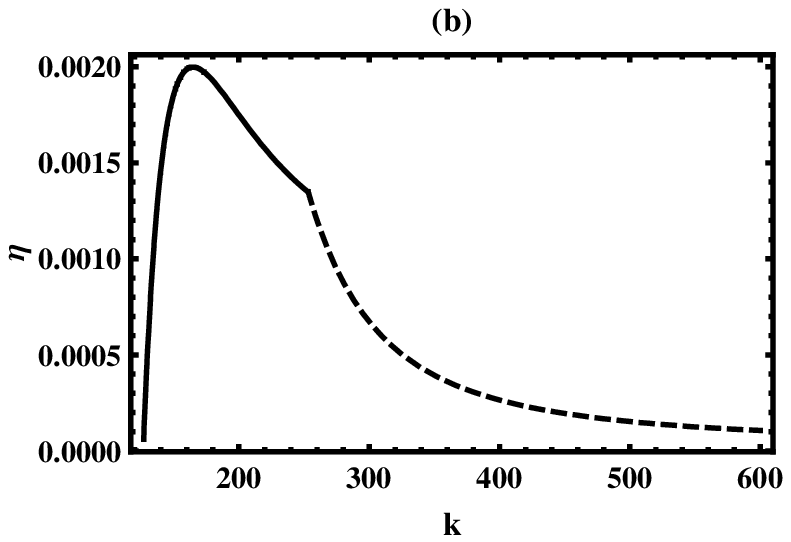}
}
\caption{ (a)  $\eta$  versus  $D$. 
The solid line represents  $k=0.8 k_c$ (stretched state), dashed line   $k=1.5k_c$ (coiled state), and dotted line  $k=\infty$ (globular state).      (b) $\eta$ versus  $k$ for   $N=50$ and $D=1.5$. In both figures we  fix $\Omega=0.01$.
}
\end{figure}

Let us now fix the polymer length $N = 50$ and vary the rescaled spring constant $k$  where $k_c=253.3$ is the critical spring constant that demarcates the polymer conformations at  the barrier.
 Figure 5a exhibits  $\eta$ versus $D$, where the curves depict 
   $k=0.6  k_c$, $k=2.5  k_c$ and $k=\infty$, respectively  representative of    stretched,  coiled, and  globular conformation at the barrier. As the spring constant (stiffness) increases, the chain shows  less resonant peaks at larger optimal noise strength. This means that larger thermal energy is required to drive the stiffer chain to move coherently with an external signal. On the other hand Figure 5b exhibits the plot of $\eta$ as a function of $k$ for $N = 50$. As $k$ becomes very small, $\eta$ tends to be very small as the whole monomers becomes  non-interacting (non-cooperative). When $k \rightarrow \infty$, $\eta$ becomes small again as globular polymer lacks the  enough  flexibility to conform to external driving.  At certain optimal value,  in between  $\eta$  attains an optimum value which is an increasing function of $D$.

In summary, we have studied the stochastic resonance of a single flexible polymer moving in a bistable potential, or flowing within an equivalent fluidic channel, both of which modulate over a mesoscopic or even a macroscopic scale. At an optimal noise strength, the dynamics of the chain, otherwise random, shows a SR, i.e., it moves in coherence and resonance with a periodic driving force. Even with the noise strength (temperature) fixed, the polymer display a novel kind of entropic SR by  responding  cooperatively to external driving in a most efficient way at  optimal chain lengths and elastic constants, owing to chain flexibility and conformational transition. Utilizing their self-organizing behaviors, we may learn bio-molecular machineries of living, and clever ways of manipulating them such as efficient separation methods.


\begin{thebibliography}{47}
\bibitem{b4} L.  Gammaitoni, P. H\"anggi, P. Jung and F. Marchesoni, Rev. Mod. Phys. {\bf 70}, 223 (1998).
\bibitem{b5} A. Bulsara,  L.  Gammaitoni, Physics Today  {\bf 49}, 39 (1996).
\bibitem{b3} R. Benzi, G. Parisi, A. Sutera and A. Vulpiani, Tellus  {\bf 34}, 10 (1982).
\bibitem{bb3} A Palonpon ,  J. Amistoso, J. Holdsworth,  W.Garcia and  C. Saloma,  Optics Letters {\bf 23},  1480 (1998).
\bibitem{b7}  A. A Priplata, B.L Patritti, J.B Niemi, $et$ $al$,  Ann Neurol.  {\bf 59}, 4 (2006).
\bibitem{b6}  K. Wiesenfeld, F. Moss,  Nature {\bf 373}, 33 (1995).
\bibitem{b9} P. S. Burada, G. Schmid, D. Reguera, M. H. Vainstein, J. M. Rubi, and P. H\"anggi, Phys. Rev. Lett. {\bf 101}, 130602 (2008).
\bibitem{b10}  I. Goychuk and P. H\"anggi, Phys. Rev. Lett. {\bf 91}, 070601 (2003).
\bibitem{b11} Y. W. Parc, D. Koh, W. Sung, Eur. Phys. J. B, {\bf 69}, 127 (2009).
\bibitem{a1} R. Zwanzig, J. Phys. Chem. {\bf 96},  (1992);
G. W. Slater, H. L. Guo, and G.I. Nixon, Phys. Rev. Lett. {\bf 27}, 1170 ( 1997);
P. S. Burda,G. Schmid, P. Talkner, P. H\"anggi, D. Reguera, J.M.Rubi, BioSystem, 93 (2008) 16.
\bibitem{a2}J. Li,  D. Stein,  C. McMullan,  D. Branton,  M. J. Aziz, and J. A. Golovchenko,   Nature  {\bf  412}, 166  (2001);
J. Storm, J. H. Chen,  X. S. Ling,  H. W. Zandbergen, and C. Dekker, Nature Mater, {\bf 2}, 537 (2003); C.  Dekker,  Nat. Nanotechnol. {\bf 2}, 209  (2007); K. Healy,  B. Schiedt, and A. P. Morrison,  Nanomedicine  {\bf 2}, 875 (2007).
\bibitem{b12} P.J. Park and W. Sung, J. Chem. Phys. {\bf 111}, 5259 (1999).
\bibitem{e2} S. Lee and W. Sung, Phys. Rev. E {\bf 63}, 021115 (2001).
\bibitem{a3} T. Ikonnen, T. Ala-Nissila and W. Sung, manuscript in preparation.
\bibitem{b15}  S.B. Smith, L. Finzi, and C. Bustamante,  Science {\bf 258}, 1122 (1992).
\bibitem{b16}K.L. Sebastian and Alok K.R. Paul, Phys. Rev. E {\bf 62}, 927 (2000).
\bibitem{b17}  J. F. Lindner, B. K. Meadows, W. L. Ditto, M. E. Inchiosa, and A. R. Bulsara, Phys. Rev. Lett. {\bf 75}, 3 (1995); Phys. Rev. E {\bf 53}, 2081 (1996).
\bibitem{b18} F. Marchesoni, L. Gammaitoni, and A. R. Bulsara, Phys. Rev. Lett. {\bf 76}, 2609 (1996). 
\bibitem{b20} R. Kubo. Rep. Prog. Phys. {\bf 29},  255 (1966); H.B. Callen and T.A. Welton. Phys. Rev. {\bf 83} 34 (1951). 
\end{thebibliography}
\end{document}